\pdfoutput=1
\documentclass{article}
\usepackage[english]{babel}
\usepackage[T1]{fontenc}
\usepackage[utf8]{inputenc}
\usepackage{amsmath}
\usepackage{mathptmx}
\usepackage{amssymb}
\usepackage[hyphens]{url}
\usepackage{hyperref}  
\usepackage{longtable}
\usepackage[normalem]{ulem}
\usepackage{multirow}
\usepackage{indentfirst}
\usepackage[algoruled,linesnumbered, longend, noline]{algorithm2e}
\usepackage{graphicx}
\providecommand{\keywords}[1]{{\textbf{Keywords--} #1}}
\begin{document}

\title{Handling Flash-Crowd Events to Improve the Performance of Web Applications}

\author{
  Ubiratam de Paula Junior\\
  L\'ucia M. A. Drummond \\
  \normalsize{Institute of Computing}\\
  \normalsize{Fluminense Federal University}\\
  \normalsize{\texttt{\{upaula,lucia\}@ic.uff.br}}
  \and
  Daniel de Oliveira\\
  Yuri Frota\\
  \normalsize{Institute of Computing}\\
  \normalsize{Fluminense Federal University}\\
  \normalsize{\texttt{\{danielcmo,yuri\}@ic.uff.br}}
  \and
  Valmir C. Barbosa\\
  \normalsize{COPPE}\\
  \normalsize{Federal University of Rio de Janeiro}\\
  \normalsize{\texttt{valmir@cos.ufrj.br}}
}
\date{\textit{Submitted to the 30th Symposium On Applied Computing (2015).}}

\maketitle

\begin{abstract}
Cloud computing can offer a set of computing resources according to users' demand.  It is suitable to be used to handle flash-crowd events in Web applications due to its elasticity and on-demand characteristics. Thus, when Web applications need more computing or storage capacity, they just instantiate new resources. However, providers have to estimate the amount of resources to instantiate to handle with the flash-crowd event. This estimation is far from trivial since each cloud environment provides several kinds  of  heterogeneous resources, each one with its own characteristics such as bandwidth, CPU, memory and financial cost. 
In this paper, the Flash Crowd Handling Problem (FCHP) is precisely defined  and formulated as an integer programming problem.   A new algorithm  for handling with  a  flash crowd  named FCHP-ILS is also proposed.  With FCHP-ILS the Web applications can replicate contents in the already instantiated resources and  define the types  and amount of resources  to instantiate in the cloud during a flash crowd.  Our approach is evaluated considering real flash crowd  traces obtained from the related literature. We also present a case study, based on a  synthetic dataset representing  flash-crowd events in  small scenarios aiming at the comparison  of  the proposed approach against Amazon's Auto-Scale mechanism.
\end{abstract}

\begin{center}
\end{center}

\section{Introduction}\label{sec:introduction}

The number of accesses to online contents on the Internet has been growing steadily in the last few years. This growth is mainly caused by several factors, \textit{e.g.} the emergence of high speed networks and the diffusion of mobile devices such as phones and tablets, which increased ubiquity. Therefore, any person, anywhere in the world with a mobile device on hands, can search for content and download it from the Internet. This large and ever growing number of accesses becomes an important issue for service providers since they have to meet QoS requirements such as availability and download speed. Besides the aforementioned increasing number of potential users, service providers also have to consider the variation in the popularity of certain contents over time. This variation of popularity is currently fostered by social networks such as Facebook and Twitter because messages post in these environments can turn a specific content into a viral. For example, during the Oscars 2014 ceremony, Ellen DeGeneres twitted a "selfie" with several top celebrities \cite{selfie}. Ellen's tweet reached incredible numbers (in less than 24 hours) such as: (i) 8.1 million people saw the tweet a total of 26 million times; (ii) 13,711 Web pages embedded the tweet, and those embeds were seen 6.8 million times; (iii) it was seen 32.8 million times and (iv) it was retweeted more than 3.2 million times. This number of visualizations and accesses made Twitter page to be unable to be accessed by users for several minutes during Sunday night. This type of increase in popularity of an online content in a short period of time is named a flash-crowd event \cite{artigo-10}. 
In flash-crowd events, all requests are legitimate, \textit{i.e.} the users \textit{really want} to access the specific content. However, since the number of accesses increases in a short period of time, this can be a problem. Service providers may not be ready to increase the number of Web servers to deploy these resources and the final effect is commonly a reduction in QoS. In more critical cases (such as 11th September attacks to WTC), the continuous accesses to these Web servers can produce a complete (and undesired) halt.  Thus, it is required that flash-crowd events should be detected and mitigated as early as possible to avoid future problems in the Web servers. This mitigation is commonly an increase of the number of Web servers that store contents and it can be fostered by the use of clouds \cite{Vaquero2008}.

Cloud computing can offer a set of computing resources according to users' demand. These resources can be instantiated on demand and the user can choose only to pay during the period of time they use the resource or a fixed price during a period of time (monthly, for example). Thus, cloud computing model is suitable to be used to handle flash-crowd events due to its elasticity and on-demand characteristics. This way, when service providers need more computing capacity, they just instantiate new resources. On the other hand, when the flash-crowd events ends, service providers can destroy these resources. However, providers have to estimate the amount of resources to instantiate to handle with the flash-crowd event. This estimation is far from trivial since each cloud environment provides several kinds  of  heterogeneous resources, each one with its own characteristics such as bandwidth, CPU, memory and financial cost.  Also,  this estimation has to be performed quickly and without under and overestimations. Under and overestimations can either slowdown the access to the content or produce high financial costs, respectively.

In this paper, the Flash Crowd Handling Problem (FCHP) is precisely defined  and formulated as an integer programming problem. A new algorithm for handling with  a  flash crowd  named FCHP-ILS is also proposed. FCHP-ILS is  based on a metaheuristic called ILS (\textit{Iterated Local Search}). With FCHP-ILS the service provider can define the types of resources (\textit{i.e.} Web servers) to instantiate in the cloud during a flash crowd. This estimation uses  as input  the  several types of  resources, characterized by their total storage and bandwidth, available in the cloud and client requests. We compare the FCHP-ILS with the exact solution given by the proposed model, by using as input flash crowd traces obtained  from the related literature. We also present a case study, based on a  synthetic dataset representing  flash-crowd events in small scenarios, to show that FCHP-ILS is capable of estimating the amount of resources to instantiate while improving resource utilization and financial costs. This experimental evaluation was conducted aiming at the comparison  of  the proposed approach against Amazon's Auto-Scale mechanism \cite{AS}.
\section{Related Work}
\label{rel_work}

Several  papers in the literature tackle the flash crowd problem. Actually, Jung,  Krishnamurthy and Rabinovich \cite{artigo-10} provide a complete study about flash crowd problem thus allowing researchers to create new strategies for Web sites to quickly discard malicious requests. While many of these solutions are focused on applications that use Content Distribution  Networks (CDN), some can also be applied generally to any type of Web applications. Additionally, we claim that  the possibility of instantiating new resources on demand using  clouds can foster the development of new solutions to handle with flash crowds. Following, we present some of the most related approaches.

Broberg \textit{et al.} \cite{metacdn} propose a general purpose framework called MetaCDN, which interacts with cloud providers to implement an overlay network that can be used as a CDN in the cloud. MetaCDN ease the task of consumers to harness the performance and coverage of numerous  "Storage Clouds" since it provides a single namespace to allow for integration in a transparent way.

Chen and Heidemann \cite{Chen} propose an adaptive admission control mechanism named NEWS that aims at protecting networks from flash crowds and maintain high performance for users. One advantage of NEWS is that it is able to detect flash crowds based on performance degradations and then start to mitigate the flash crowd. In the presented experiments, NEWS was able to detect flash crowds in 20 seconds. After that, it re-dimensions the network to maintain the high performance. 

Pan \textit{et al.} \cite{fcan2,fcan} in their paper propose FCAN. FCAN is an approach that implements a P2P overlay over the real network to distribute the flash traffic from origin Web server. It is based on DNS redirection to route the requests in a balanced way,  however, it does not increase or decrease the amount of resources during the flash crowd.

Tian, Fang and Yum \cite{tian} propose a defense system to react to flash crowds in Web services. Their systems try to employ dynamic bandwidth using a method based on the Vickrey auction. Their idea is that the availability of Web services is improved while the utility and availability are maximized. Their system was evaluated using simulated environments.

Stavrou \textit{et al.} \cite{artigo-1} propose a system called PROOFS that implements a P2P overlay which allows clients, that seek popular contents, to obtain them from other clients. PROOFS was only evaluated using simulated environments.

Moore \textit{et al.} \cite{Moore} propose an elasticity management framework that consider as input a series of reactive rule-based strategies and generates a proactive strategy as outcome. They combine reactive and predictive auto-scaling techniques, \textit{i.e.} they try to predict when a flash crowd (they call as peaks) will occur.

Vasar \textit{et al.} \cite{Vasar} propose a framework that integrates a set of monitoring tools. The framework is designed to aid users to test applications under various configurations and workloads. The proposed framework supports dynamic server allocation based on incoming load using a response-time-aware heuristic. In their paper they compared the proposed approach against Amazon Auto Scaling mechanism.

Tang \textit{et al.} \cite{tang2014} propose a systematic framework for dynamic request allocation and service capacity scaling in a cloud-centric media network. They provide simulations that suggest that their proposed dynamic allocation and service capacity scaling mechanism outperforms other existing allocation methods.

There are also some commercial solutions such as the combination of Amazon’s Auto Scaling and Load Balancing mechanisms \cite{AS,ELB}. Amazon’s Auto Scaling mechanism allows for users to horizontally scale the amount of virtual machines according to current environment status, \textit{i.e.} during a flash crowd the number of virtual machines the user has can be increased to maintain performance, and decreases automatically during demand lulls to minimize financial costs. Flash crowds are identified in Amazon’s environment using the cloud watch mechanism that allows for identifying when the income traffic trespassed a pre-defined threshold. General Web applications are suitable to benefit from Auto Scaling mechanism. However, Auto Scaling only creates the virtual machines. To provide load balancing according the existing virtual machines we have to use the Load Balancing mechanism. The Amazon’s Elastic Load Balancing (ELB) automatically distributes incoming requests over the multiple instantiated virtual machines.
However, providing these kind of scaling rules in commercial solutions is difficult, error-prone and asks some infrastructure expertise. In this way, Kouki and Dedoux \cite{SCAling} propose SCAling, a platform and an approach driven by Service Level Agreement (a formal contract between a service provider and a service consumer on an expected QoS level) requirements for Cloud auto-scaling.

Although there are several related papers in the literature, to the best of our knowledge none of them proposes a mathematical treatment for the problem, considering different characteristics of the real problem jointly. 
\section{Traces of Flash Crowds}
\label{sec:background}
	
	Flash crowds can be characterized by several factors, such as duration, growth rate, and locality. A flash-crowd event at a web site has three main phases \cite{artigo-2}: a \textit{ramp-up} phase, a \textit{sustained-traffic} phase, and a \textit{ramp-down} phase. Starting at a traffic pattern that is considered normal, during the ramp-up phase traffic rises significantly for a small number of contents and can stay high for some time, depending on the kind of flash crowd. In the ramp-down phase the number of accesses gradually decreases until traffic becomes once again normal. Due to the difficulty in obtaining real traces to evaluate the method for handling flash crowds, a generator of synthetic flash crowds was developed. This generator can produce several kinds of traces by modeling flash crowd's main phases and taking into account some of the characteristics of real flash crowds. It is based on sampling from a variety of beta distributions, as discussed next.
		
	Assume any given content and let $X_t$ be a discrete random variable representing the number of accesses to that content at time $t$. Assume further that $X_t$ takes values from the domain $[0,U]$ for some integer $U>0$. Our goal is to devise a probability mass function on the integers in $[0,U]$ such that sampling from the corresponding distribution can generate a number of accesses to the content in question for any time $t$. Although our random variable is not continuous, the use of a bounded support such as $[0,U]$ immediately suggests the beta density function, whose support is $[0,1]$. What we do is to adopt this function nevertheless, noting that every sample is to be rounded off to an integer after being scaled up by the factor $U$. The beta density function to be used in sampling a value $x$ for $X_t$, henceforth denoted by $p_t(x)$, is given by
$$
p_t(x)=\frac{x^{\alpha_t-1}(1-x)^{\beta_t-1}}{\Gamma(\alpha_t)\Gamma(\beta_t)/\Gamma(\alpha_t+\beta_t)},
$$
where $\alpha_t$ and $\beta_t$ are both positive and control the function's shape through its mean, $\alpha_t/(\alpha_t+\beta_t)$, and variance, $\alpha_t\beta_t/(\alpha_t+\beta_t)^2(\alpha_t+\beta_t+1)$.

	This shape parameterization is convenient for the generation of flash crowd accesses, particularly during the ramp-up and ramp-down phases. To see this, consider two instants $t_0<t_1$ and the modeling of the ramp-up phase of a flash crowd between $t=t_0$ and $t=t_1$. We first choose $\alpha_{t_0}\ll\beta_{t_0}$, so that $X_{t_0}$, the number of accesses right before the ramp-up phase, has a low mean and a relatively low variance. Likewise, in order to get high values for $X_{t_1}$, that is, right past the ramp-up phase, we choose $\alpha_{t_1}\gg\beta_{t_1}$. The latter can be achieved, for example, by setting $\alpha_{t_1}=\beta_{t_0}$ and $\beta_{t_1}=\alpha_{t_0}$, which implies a high mean and also both $\alpha_{t_1}+\beta_{t_1}=\alpha_{t_0}+\beta_{t_0}$ and $\alpha_{t_1}\beta_{t_1}=\alpha_{t_0}\beta_{t_0}$, thence the same relatively low variance as before. This particular choice for $\alpha_{t_1}$ and $\beta_{t_1}$ is therefore quite handy, because it immediately allows all intermediate values of $t$ (i.e., the ramp-up phase itself) to be handled easily: We simply let
$$
\alpha_t=y_t\alpha_{t_0}+(1-y_t)\beta_{t_0}
$$
with $0\le y_t\le 1$ and
$$
\beta_t=\alpha_{t_0}+\beta_{t_0}-\alpha_t.
$$
That is, both $\alpha_t$ and $\beta_t$ are convex combinations of $\alpha_{t_0}$ and $\beta_{t_0}$. Setting $y_t=1$ recovers $p_{t_0}(x)$; setting $y_t=0$ recovers $p_{t_1}(x)$.

	As for sampling $X_t$ during the ramp-up phase itself, all that is left to do is to decide the rate at which $y_t$ is decreased from $1$ at $t=t_0$ to $0$ at $t=t_1$. Here we let $y_t$ decrease in such a way that $1-y_t$ increases exponentially. That is,
$$
y_t=\frac{e^{\gamma(t_1-t_0)}-e^{\gamma(t-t_0)}}{e^{\gamma(t_1-t_0)}-1}
$$
for some $\gamma>0$, which implies a ``generally convex'' growth pattern in the number of accesses. The number of accesses for any time prior to the ramp-up phase can be sampled as $X_{t_0}$. For times in the sustained-traffic phase, the number of accesses can be sampled as $X_{t_1}$.

	Modeling instead a ramp-down phase from $t_0$ to $t_1$ can be done essentially along the same lines. We simply let both $\alpha_t$ and $\beta_t$ be given as above and increase $y_t$ from $0$ at $t=t_0$ to $1$ at $t=t_1$. We do this in such a way that $y_t$ increases exponentially, that is,
$$
y_t=\frac{e^{\gamma(t-t_0)}-1}{e^{\gamma(t_1-t_0)}-1}
$$
with $\gamma>0$, which promotes a ``generally concave'' decay pattern in the number of accesses. The number of accesses for any time past the ramp-down phase can be sampled as $X_{t_1}$.

\section{Problem Definition}
\label{sec:formulation}

In order to avoid that a set of contents becomes unavailable during a flash crowd it is necessary to instantiate more resources to  attend  the increasing demand for  them.  There are  two possible approaches here, either to replicate  such contents in the already instantiated  Web servers or to instantiate new servers.  Replication may be not  feasible,  when there is no enough storage or bandwidth in the original  Web  servers  to attend the new demand satisfactorily. Because  flash-crowd events are unpredictable, the use of cloud servers can be a more attractive approach in those cases.
Thus, the flash crowd Handling Problem (FCHP), introduced here, can be defined as follows. Let $S=S_p \cup S_e$ be the set of all servers (virtual or physical), where $S_p$ and $S_e$ are defined as the set of Web application servers and the set of servers available for hire in the cloud, respectively.
We also consider a set of requests $R$ to be attended and a set of  contents $C$ offered by the Web application in a period of time $t\in T = \{T_1,\dots,T_{f}\}$.
Each server $j \in S$ has a storage capacity and a maximal bandwidth. Similarly, each content $k \in C$ has a start time (\textit{i.e.} the period that content $k$ is submitted in the application) and an origin server. Each request $i \in R$ requires a content and each content $k \in C$  has a size.

The FCHP is the problem of copying replicas of contents on the servers and hiring new elastic servers in order to handle the requests during the flash crowd, respecting the available storage and bandwidth and trying to minimize the  cost function defined by three sums of time: (i) the time cost $c_{i}$ to handle request $i \in R$, defined by $\sum_{i\in R}  \sum_{j \in S} \sum_{t\in T} c_{i}x_{ijt}$, where the binary variable $x_{ijt}$ indicates if the request $i$ is attended by server $j$ in period $t$, (ii) the sum of backlogging time penalty $p_{it}$ (\textit{i.e.} the penalty to postpone the attendance of request $i$ that arrived in period $t$), defined by $\sum_{i \in R} \sum_{t \in T} p_{it}b_{it}$, where $b_{it}$ indicates the postponed amount of request $i$ in period $t$, and (iii) the sum of time cost $h_{k}$ to copy the content $k$, defined by $\sum_{k \in C} \sum_{j \in S} \sum_{l \in S} \sum_{t \in T} h_{k}w_{kjlt}$, where the binary variable $w_{kjlt}$ indicates if the content $k$ is copied from server $j$ to server $l$ in period $t$.  

The described scenario can be formulated as an integer programming problem, named FCHP-IP, where the cost function is presented next:

\begin{equation*}\label{fo4}
\begin{array}{c}
	\min \displaystyle 	\sum_{i\in R}  \sum_{j \in S} \sum_{t\in T} c_{i}x_{ijt} + \sum_{r \in R} \sum_{t \in T} p_{it}b_{it} + \sum_{k \in C} \sum_{j \in S} \sum_{l \in S} \sum_{t \in T} h_{k}w_{kjlt} 
	\end{array} 
\end{equation*}

Due to space limitations, we will briefly comment the FCHP-IP constraints. The formulation must ensure that every request is completely attended by a server that has a replica of the desired content. 
Moreover, the attendance of a request must not exceed the server bandwidth, and at least one replica of each content must exist in each period of time. In a content replication, the server must have enough storage capacity to store the new replica. Finally, a cloud server can only handle requests if it has been hired.
\section{Proposed Solution}
\label{heuristic}


Exact procedures have often proved incapable of finding solutions as they are extremely time-consuming, particularly for real-world problems. Conversely, heuristics and metaheuristics provide sub-optimal solutions in a reasonable time. Moreover, the solutions produced by the proposed mathematical formulation above can not be used in practice, because it uses future knowledge about content requests.

In this context, we have designed and implemented an ILS-RVND heuristic \cite{ils}. The ILS-RVND heuristic can be defined as a multi-start method that uses (i) a random/greed heuristic in a constructive phase, (ii) a Variable Neighborhood Descent with Random neighborhood ordering (RVND) in the local search phase and (iii) perturbations moves as  a diversification mechanism. 
The main steps of the ILS-RVND are described in Algorithm \ref{alg_ils-rvnd}. The multi-start method executes $iter\_max$ iterations, where at each iteration the constructive procedure generates an initial solution $s$ (line 2) that may be improved
by the RVND method (line 3). The internal loop (lines 5-14) aims to improve the initial solution by "shaking" solution $s$ with a perturbation mechanism (line 7) and re-applying the local search method. The parameter $level\_max$ represents the maximum level of perturbation applied to the current solution. Next, we provide a short explanation of the main components of
the ILS-RVND heuristic.

\begin{algorithm}[thp]
\footnotesize
\caption{\footnotesize ILS-RVND  \label{alg_ils-rvnd}}
	\For{$i:=1$ \textbf{to} $iter\_max$}{
		$s:=constructive\_phase()$;\\
		$s:=RVND(s)$;\\
		$level:=0$;\\
			\While{$level<level\_max$}{
			$s':=s$;\\	
			$s':=perturbation(s', level)$;\\
			$s':=RVND(s')$;\\
			\eIf{$f(s')<f(s)$}{
				$s:=s'$; $level:=0$;\\
			}
			{
				$level:=level+1$;\\
			}
	}
	\If{$f(s)<f(s^*)$}{
		$s^*:=s$;\\
	}
	}
\end{algorithm}


The constructive phase consists of a greedy heuristic that builds a feasible solution by assigning each request to the cheapest server with enough bandwidth for each period of time. The chosen server must hold the required content or have enough storage to keep it. The method first tries to exhaust the set of Web application servers $(S_p)$ before start hiring servers available  in the cloud $(S_e)$. 
When there is no available server capable of attending the request, including the cloud servers that could be hired by the Web application, one of the following approaches is applied. If there is a server with enough bandwidth and no available storage  a content  is removed according to the $LRU$ (\textit{Least Recently Used}) strategy. Otherwise, the request attendance is postponed and a  backlog cost is added to the  associated cost function. The randomization of this heuristic is achieved by creating a random order of the  requests.

The local search is executed by a VND algorithm \cite{Mladenovic1997} with a random neighbourhood ordering (RVND). 
Let $N$ be an unordered set of neighbourhood structures. Whenever one neighbourhood fails to improve the current solution, the RVND randomly chooses another neighbourhood in $N$ to continue the search. The local search halts when no better solution is found in the set of neighbourhood structures of the current solution.
%
%
In order to describe the neighbourhood structures, we need some additional notation. We define
a solution $\{(k_1,j_1,r_1,t_1),$ $(k_2,j_2,r_2,t_2),...\}$ as a set of $4-$tuples $(k,j,r,t)$ representing
 that content $k$ is replicated in server $j$ to attend a set of requests $r$ on period $t$. The ILS-RVND is composed by the following five neighbourhoods:

\begin{itemize}
	\item {\bf Shift} $(k,j_a,r,t) \rightarrow (k,j_b,r,t)$: transfer one tuple from a server $j_a$ to a server $j_b$. 
	\item {\bf Swap} $(k_a,j_a,r_a,t_a)$, $(k_b,j_b,r_b,t_b)$ $\rightarrow (k_a,j_b,r_a,t_a)$, \\$ (k_b,j_a,r_b,t_b)$: one tuple from a server $j_a$ is permuted with a tuple from server $j_b$.
	\item {\bf Split} $(k,j_a,r_a,t)$ $\rightarrow$ $(k,j_b,r_b,t)$,$(k,j_c,r_c,t)$: split requests from server $j_a$ between servers $j_b$ and $j_c$, where $r_a=r_b \cup r_c$. 
	\item {\bf Merge} $(k,j_a,r_a,t), (k,j_b,r_b,t) \rightarrow (k,j_c,r_a \cup r_b,t)$: merge set of requests from servers $j_a$ and $j_b$ into a new tuple in server $j_c$.
	\item {\bf $d-$Delay} $(k,j,r,t) \rightarrow (k,j,r,t+d)$: delay a tuple in $d$ periods of time. Note that we can have a positive delay $(d > 0)$ or a negative delay $(d < 0)$. 
\end{itemize}
It is important to emphasize that only feasible movements are accomplished. 
%
%
With respect to the perturbation mechanism, we perform multiple Shift, Swap, Split and Merge movements randomly chosen in such a way that the resulting modification is sufficient to escape from local optima and analyse different regions of the search space.
The function $f$ is the same of FCHP-IP.
\section{Experimental Results}
\label{sec:result}

	In this section, we present the results for two accomplished tests. The fist one is a  comparison between the FCHP-IP mathematical formulation and the FCHP-ILS heuristic. The second is a  comparison between the FCHP-ILS heuristic and the combination of  Amazon’s Auto Scaling and Load Balancing mechanisms \cite{AS,ELB}  executed in two real small-scale scenarios.

\subsection{Comparing FCHP-IP and FCHP-ILS}
	 We evaluate FCHP-ILS, in terms of  quality of solution and execution time,  by comparing it with the solutions given by the formulation FCHP-IP when solved with the CPLEX $12.5.1$ \cite{cplex}. Those tests were run  in a computer with a processor Intel Core i7-3820 3.60GHz  and  a 32Gb memory running Ubuntu $12.04$. The  FCHP-ILS was implemented in the Programming language C/C++, gcc version $4.6.3$.
	
	 Those programs were run over a set of instances created from a  real trace obtained from the 1998 World Cup site \cite{copa98}.  Because this trace presents a huge number of requests, the total time of the trace was discretized in hours and only the requests for the ten most accessed contents in each interval of time were considered. Thus, instances with reduced size, but still presenting flash-crowd events, were created. Remark that the original instance could not be solved in a reasonable time by the CPLEX. 

	 Table \ref{Instance} presents for each of the twelve created instances, the  associated number of contents, requests and number of periods (hours). The last two instances were created by the synthetic trace generator and used for the real experiments, described in the next section.
	
\begin{table}[htp]
\small
\setlength{\tabcolsep}{3mm}
\caption{Instance description}
\begin{center}
\setlength{\tabcolsep}{0.6mm}
\begin{tabular}{|c|r|r|r|}
\hline 
Instance & \# of Contents  & \# of Requests & Periods (hours) \\
\hline
1	&	44	&	1798		&	12		\\
2	&	60	&	1696		&	12	    \\
3	&	88	&	983	    	&	12		\\
4	&	52	&	3546		&	24		\\
5	&	96	&	3427		&	24		\\
6	&	150	&	1922		&	24		\\
7   &   47  &   4327	    &   36      \\
8   &   83  &   4073        &   36      \\
9   &   99  &   1676        &   36      \\
10   &  69  &   7110 		&   48      \\
11   &   3  &   105         &   60      \\
12   &   4  &   186         &   60      \\
\hline
\end{tabular}
\end{center}
\label{Instance}
\end{table}
	
Table \ref{teorico} shows the results obtained by FCHP-IP and FCHP-ILS. The first column identifies the instance. The following six columns present the results obtained by FCHP-IP: number of hired  on demand servers, the total cost, the attendance cost, replication cost, backlog cost and the execution time to obtain the optimum solution. Following, the next columns present the same results of FCHP-ILS. Finally, the last column, shows the gap between the solutions given by FCHP-ILS and FCHP-IP. The  values shown for FCHP-ILS are averages of three executions, where  $3$, $7$ and $1$ were used for the number of iterations, the number of perturbations and the value of $d$ ({\bf $d-$Delay} neighbourhood), respectively.

\begin{table}[htp]
\scriptsize
\caption{Results of FCHP-ILS Metaheuristic and FCHP-IP Mathematical Formulation using CPLEX.}
\begin{center}
\setlength{\tabcolsep}{1mm}
\begin{tabular}{|c|crrrcr|crrrcr|c|}
\hline 
\multirow{3}{0.5cm}{Inst} & \multicolumn{6}{|c|}{FCHP-IP} & \multicolumn{6}{|c|}{FCHP-ILS} & Gap\\
\cline{2-13}
 & Serv & \multicolumn{4}{|c|}{Time Cost} & & Serv & \multicolumn{4}{|c|}{Time Cost} & & \\
 &  OD & \multicolumn{1}{|c}{Total} & Attend & Repli & \multicolumn{1}{c|}{Back} & Time & OD & \multicolumn{1}{|c}{Total} & Attend & Repli & \multicolumn{1}{c|}{Back} & Time & (\%)\\
\hline 
1	&	14	&	1726.9	&	1639.6	&	87.3	&	0	&	134.6	&	7	&	2647.0	&	2486.3	&	160.8	&	0	&	93.7	&	53.3	\\
2	&	8	&	1993.6	&	1869.2	&	124.4	&	0	&	174.9	&	6	&	2139.6	&	1979.2	&	160.4	&	0	&	57.8	&	7.3	\\
3	&	8	&	607.4	&	446.6	&	160.8	&	0	&	44.2	&	2	&	749.4	&	573.3	&	176.1	&	0	&	73.8	&	23.4	\\
4	&	8	&	3512.8	&	3349.2	&	163.6	&	0	&	546.5	&	15	&	5570.4	&	5242.5	&	327.8	&	0	&	359.1	&	58.6	\\
5	&	26	&	2065.3	&	1775.4	&	289.9	&	0	&	230.1	&	5	&	2390.8	&	2065.4	&	325.4	&	0	&	168.5	&	15.8	\\
6	&	55	&	7167.0	&	6664.4	&	502.6	&	0	&	36619.4	&	9	&	8355.6	&	7604.4	&	751.2	&	0	&	185.7	&	16.6	\\
7	&	12	&	4991.3	&	4785.4	&	205.9	&	0	&	781.4	&	19	&	7914.9	&	7505.4	&	409.5	&	0	&	319.6	&	58.6	\\
8	&	10	&	3721.1	&	3344.6	&	376.5	&	0	&	610.2	&	15	&	5616.8	&	5084.6	&	532.2	&	0	&	230.4	&	50.9	\\
9	&	19	&	1179.1	&	750.4	&	428.7	&	0	&	280.2	&	8	&	1622.2	&	1040.4	&	581.8	&	0	&	79.7	&	37.6	\\
10*	&	43	&	7586.3	&	7202.0	&	384.3	&	0	&	2483.8	&	26	&	11763.0	&	11087.0	&	681.7	&	0	&	1308.6	&	55.1	\\
11*	&	440	&	28451.7	&	21957.4	&	6494.3	&	0	&	7388.8	&	123	&	28764.8	&	21957.4	&	6807.4	&	0	&	967.5	&	1.1	\\
12*	&	239	&	41228.2	&	32208.9	&	9019.3	&	0	&	14382.2	&	282	&	42036.2	&	32208.9	&	9827.3	&	0	&	6564.4	&	2.0	\\
\hline
Avg	&		&		&		&		&		&	5306.4 	&		&		&		&		&		&	867.4 	&	31.7	\\
\hline 
\end{tabular}
\end{center}
\label{teorico}
\end{table}

	In this table, we can observe a high gap for some instances. This is due to the difficult of the considered problem. The FCHP problem extends the Replica Placement Problem (RPP), which belongs to the NP-hard class \cite{aioffi05}. For small instances of the problem,  FCHP-ILS   takes a long time to find good solutions. However, as the input data increases, the FCHP-IP is not capable of proving the optimality of the solution in a reasonable time. Moreover,  FCHP-IP needs a higher memory capacity to solve bigger instances.  The results for instances $6$, $11$ and $12$ show that FCHP-ILS is more suitable in those cases, when  the gap and the execution time were reduced when compared with the averages results. Furthermore, in the last three instances, the CPLEX could not prove the optimality of the solution due to lack of memory. Remark that the two last instances were used in the real experiments.
	
\subsection{Comparing FCHP-ILS and Auto Scaling in real scenarios}	
		
	In order to analyze the performance of  FCHP-ILS  in a  commercial cloud, we executed it in two smalls scenarios of the problem and compared with the mechanism provided by the  popular  Auto Scaling and Load Balancing of the Amazon.
	We used the synthetic trace generator of flash-crowd events,  described in Section \ref{sec:background}, to generate the  clients'  requests.  The  Web application was developed using Apache Tomcat $6.0$ \cite{tomcat}.
	
\begin{figure}[ht]
\centering
\begin{center}
\includegraphics[scale=0.5]{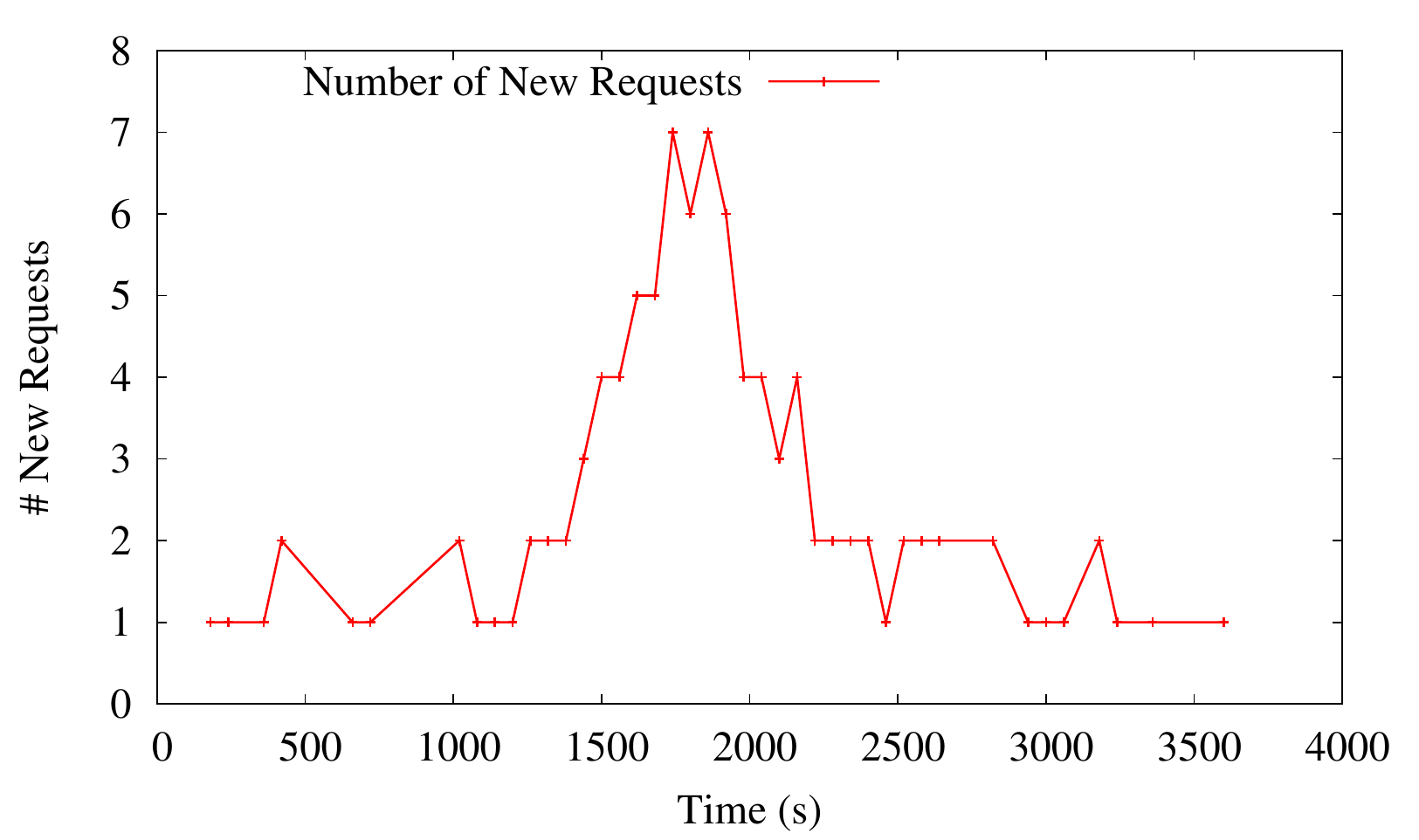} 
\end{center}
\caption{New generated requests per second for the first scenario.}
\label{fc-figure1}
\end{figure}

	In the first scenario, the Web application offers  three different contents  with  the following sizes: $1.9$, $1.5$ and $0.9$ Gbytes. The \textit{ramp-up phase} occurs between $1140$ and $1740$ seconds, the \textit{sustained traffic phase} from $1800$ to $1860$ seconds,  and the \textit{ramp-down phase} in the interval from $1920$ to $2460$ seconds, as shown in  Figure \ref{fc-figure1}. 
	
\begin{figure}[ht]
\centering
\begin{center}
\includegraphics[scale=0.5]{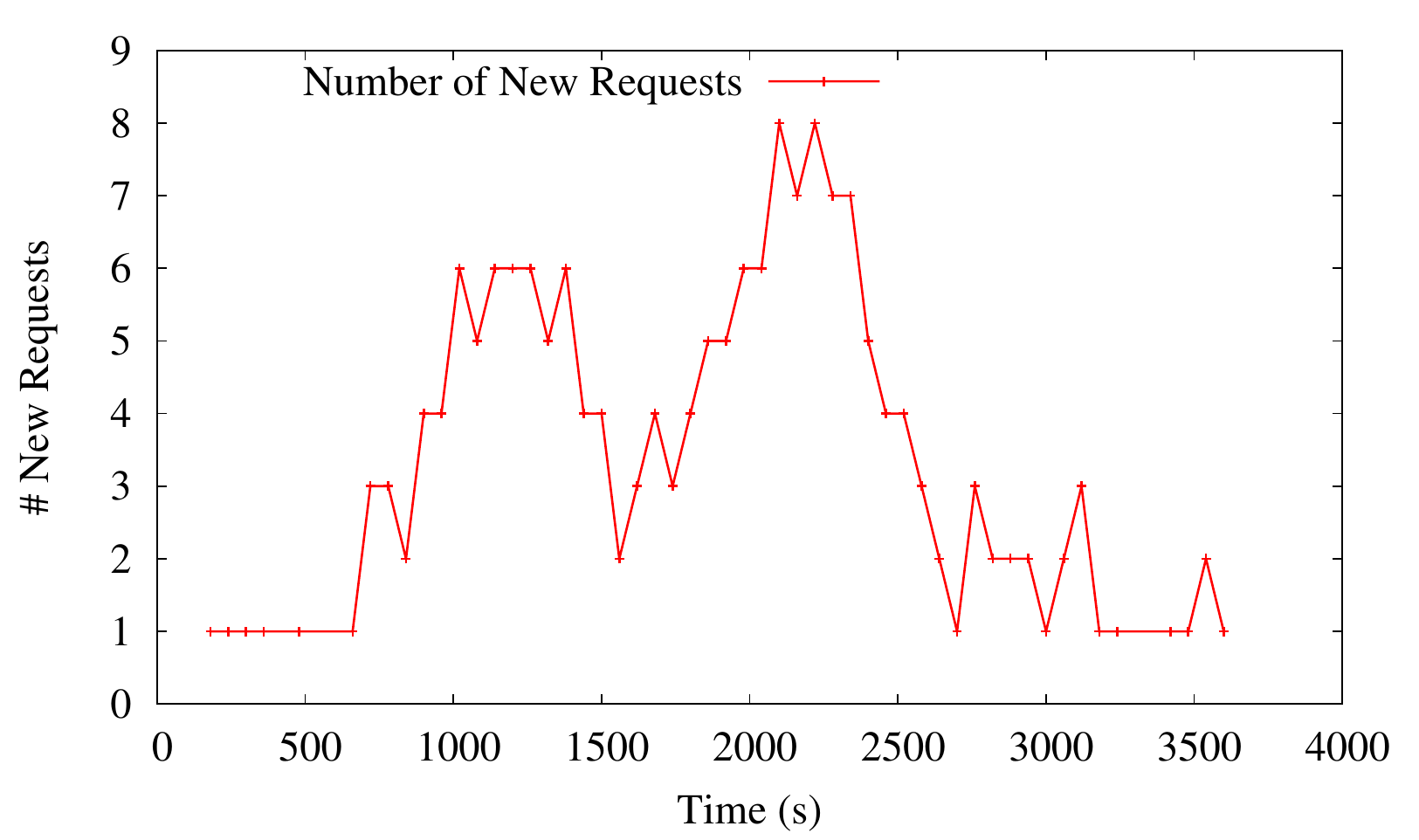} \\
\end{center}
\caption{New generated requests per second for the second scenario.}
\label{fc-figure2}
\end{figure}

	In the second scenario, the Web application offers four different contents  with  the following sizes: $0.8$, $1.0$, $1.5$ and $2.0$ Gbytes. In this scenario there are two flash-crowd events. For the first flash crowd, the \textit{ramp-up phase} occurs between $540$ and $1140$ seconds, the \textit{sustained traffic phase} from $1200$ to $1320$ seconds,  and the \textit{ramp-down phase} in the interval from $1380$ to $1860$ seconds. For the second one,  the \textit{ramp-up phase} occurs between $1500$ and $2100$ seconds, the \textit{sustained traffic phase} from $2160$ to $2220$ seconds,  and the \textit{ramp-down phase} in the interval from $2280$ to $2820$ seconds. The accesses for this scenario are shown in Figure \ref{fc-figure2}.

	In both tests, the servers' bandwidth is  $10$ Mbytes/s while  the clients' bandwidth  is $1$ Mbytes/s  bandwidth  in average. The content of $0.9$ Gbytes was the only one involved in the flash crowd for the first test and the contents of $0.8$ and $1.0$ Gbytes for the second one.
	
	The main objective to create and execute the second scenario was to analyze how these two approaches, Auto Scaling and  FCHP-ILS,  behave in a flash-crowd event with two distinct contents at overlapping periods of time. 

	To have a fair comparison, in both approaches, the flash crowd was detected in the same time and the Web application started by using two virtual machines of type \textit{m3.large}  of the Amazon, each one containing all the contents.	
	
	The Auto Scaling was configured to allocate new virtual machines (Web servers) in the beginning of the flash crowd. 
	Note that, during a flash crowd, the solution provided by Auto Scaling allocates virtual machines based on a static virtual image that must contain all contents of the Web application.
	 So it continued to use  \textit{m3.large} virtual machines,  which have  enough capacity to store the three contents. The  Load Balancing service is employed to evenly distribute the client's requests among the used  servers. 
	
	In our approach, requests are treated by a machine that has a similar role  of the one which executes the Auto Scaling and  Load Balancing service. It is responsible for directing the received requests to the virtual machines; when a flash crowd  occurs,  executing   FCHP-ILS;   and,  in accordance with the results given by FCHP-ILS, replicating contents or allocating  new virtual machines. Remark, however, that in our approach the new  allocated virtual machines can be different from the originally allocated ones. It allows for the Web application to use heterogeneous resources of the clouds. In our tests, cheaper and smaller virtual machines were allocated to store only the contents that were involved in the flash-crowd event. 
	
	In both approaches, more six virtual machines were hired during the flash crowd in the first scenario and more eight for the second one. However,  FCHP-ILS hired \textit{m3.medium} virtual machines with  enough storage capacity to keep only the contents involved in the flash crowd, which reduced the financial costs. In the first scenario, by using Auto Scaling the total financial cost was \$$4.25$, while in our approach was \$$3.11$. The execution times were similar, around $5560$ seconds. In the second scenario, the total financial cost was \$$5.60$ for the Auto Scaling solution and was \$$4.08$ for our solution. Again, the execution times were similar, around $5140$ seconds. Note that for both scenarios, with only one content involved in the flash crowd and with distinct contents at overlapping periods of time, the proposed solution, FCHP-ILS, presents satisfactory results in real environments.
\section{Concluding remarks}
\label{sec:conc}

	Our results showed that FCHP-ILS  is   efficient, when compared with the exact method and when tested in the Amazon against the Auto Scaling method.
	The proposed approach could solve the formulated problem satisfactorily, reducing the  amount of instantiated resources of the cloud during a flash crowd. 	We intend to continue the tests in commercial clouds, considering different scenarios. Moreover, we are working on a  procedure to detect flash crowds more efficiently, to aid the decisions  about content replication and new virtual machines hiring.

\bibliographystyle{plain}
\bibliography{ref}\label{sec:refs}  

\end{document}